\documentclass[sigconf,nonacm]{acmart}
\renewcommand\footnotetextcopyrightpermission[1]{} %
\AtBeginDocument{%
  }
\makeatletter
\renewcommand\@formatdoi[1]{\ignorespaces}
\makeatother

\setcopyright{none}

\begin{document}

\title{A Retrospective on Ultrasound Mid-Air Haptics in HCI}

\author{Arthur Fleig}
\affiliation{%
	\institution{Center for Scalable Data Analytics and Artificial Intelligence (ScaDS.AI) Dresden/Leipzig, Leipzig University}
	\city{Leipzig}
	\country{Germany}
}
\email{arthur.fleig@uni-leipzig.de}

\begin{abstract}
	In 2013, the UltraHaptics system demonstrated that focused ultrasound could generate perceivable mid-air tactile sensations, building on earlier explorations of airborne ultrasound as a haptic medium. 
	These contributions established ultrasound mid-air haptics (UMH) as a viable interaction modality and laid the technical and perceptual foundations for subsequent advances in Human-Computer Interaction (HCI). 
	In this extended abstract, we revisit this formative work, trace the research and design trajectories it enabled, and reflect on how UMH has supported multisensory interaction, immersion, and inclusion. 
	We also highlight how this line of research exemplifies the value of interdisciplinary collaboration to advance novel interactive technologies.
\end{abstract}

\keywords{Mid-Air haptics, acoustic radiation force, simulating touch}

\maketitle

\section{Introduction}
Haptic feedback has long been integral to interactive systems, for instance via force feedback in joysticks, tactile feedback when tapping on a smartphone screen, or vibration in gloves to provide a sense of touch in virtual reality~\cite{carlos21surveymobilear}. 
\emph{Mid-air} haptics decouples the perception of touch and actually touching an object, e.g., through air-jets~\cite{suzuki05airjet}, using lasers~\cite{lee16laser}, or ultrasound.
The latter uses the \emph{acoustic radiation force} emitted from transducers onto a user's bare hands, for which \citeauthor{iwamoto08haptics} provided the first idea and prototype~\cite{iwamoto08haptics,hoshi10tactiledisplay}. 
Based on this concept, in 2013 and 2014, a team from the University of Bristol helped pave the way for UMH to be broadly applicable within HCI -- with research papers~\cite{carter13ultrahaptics,long14volumetric,wilson14perception} (each between 140 and 760 citations according to Google scholar\footnote{https://scholar.google.com}) and commercially available products\footnote{https://www.ultraleap.com}. 
Subsequently, we revisit and reflect on this foundational research. %

\section{From Feasibility to User Experience}
The UltraHaptics system~\cite{carter13ultrahaptics} demonstrated that focused ultrasound can generate up to five concurrent perceivable tactile points above a perforated display surface. 
At that time (2013), the creation of multiple focal points was a challenge and required an algorithmic solution to drive the phase and amplitude of each transducer in a 16x20 transducer array without creating (too much) unwanted interference. 
Similarly, it was unclear to what extent users can distinguish these points and their different tactile properties. 
By combining technical contributions with user research, \citeauthor{carter13ultrahaptics} showed that users could discriminate between multiple points, especially when various points had different frequencies or after some training~\cite{carter13ultrahaptics}.
In this manner, \citeauthor{carter13ultrahaptics} established perceptual and engineering parameters for ultrasound mid-air haptic feedback. 
In a time where computing moved from physical controls to increasingly intangible interfaces such as smooth touchscreens and gestural interactions, UltraHaptics offered a compelling solution to return \emph{touch} to touchless interaction. 

Building on this foundation, in 2014 \citeauthor{wilson14perception} advanced the perceptual understanding of ultrasonic haptic feedback by exploring users' (N=14) localisation accuracy and motion perception across the hand when both hand and array do not move~\cite{wilson14perception}. 
Their experiments quantified perceptual thresholds, revealing that users could localize static points with an average error of 8.5mm. 
Moreover, the illusion of movement was stronger for larger distances, more intervening stimuli along the path, and longer duration of the stimuli. 
From these insights, \citeauthor{wilson14perception} derived design guidelines for interface designers working on haptic feedback in HCI~\cite{wilson14perception}. 

Later that same year, \citeauthor{long14volumetric} made the leap to the third dimension, and demonstrated that a large number of discrete focal points could be algorithmically composed into \emph{volumetric haptic shapes} that one could \emph{feel}~\cite{long14volumetric}. 
This required not only detecting when hands interact with (invisible) shape boundaries, but also a fast and robust algorithm for interactively synthesizing acoustic fields. 
To this end, the authors borrowed from linear algebra and calculus. 
Reformulating the problem as an (iterative) solution to a linear equation and utilizing parallelization yielded a system being able to render shapes such as spheres, pyramids, and cones in real-time. 
A small user study (N=6) showed that participants were generally able to correctly identify a shape out of five options~\cite{long14volumetric}. 
With this, UMH transitioned from isolated points of sensation to spatially continuous, volumetric touch.

\section{Ongoing Relevance in Today's HCI Research Landscape}
As applications, \citeauthor{long14volumetric} envisioned multisensory design (a car dashboard with a touch and haptic interface), mixed reality, and \emph{feeling} inaccessible objects, such as those in museum cases~\cite{long14volumetric}. 
Fast forward a decade, the principles pioneered in the foundational studies resonate strongly within HCI, and all three envisioned applications are being tackled. 

In the automotive context, UMH is used to augment the audio-visual channels and improve driving safety by reducing drivers' eyes-off-the-road time when interacting with in-car user interfaces~\cite{harrington18automotiveui,young20cardesign,korres20midair,Montano24knuckles}. 
In mixed reality, this technology is used to improve immersion~\cite{frutos19vrultrasound,pinto20xr,howard2022immersive}, enable haptic feedback for virtual music instruments, rhythm games, and typing~\cite{georgiou18vrrhythm,hwang17vrpiano,park25ultraboard}, and investigate its role on agency~\cite{evangelou2025investigating}. 
Feeling inaccessible objects include the envisioned museum case~\cite{vi17museum}, experiencing internal human organs~\cite{romanus19xrmidair}, and science communication~\cite{hajas20sciencecommunication}.
Additionally, volumetric displays emerged~\cite{marzo15holographic,paneva22optitrap,fushimi2024hologram}, moving towards Ivan Sutherland's vision of the Ultimate Display~\cite{sutherland1965ultimate}. 

UMH also aids inclusion, accessibility, and rehabilitation. 
For instance, \citeauthor{obrist15emotions} explore mediating emotions~\cite{obrist15emotions} through UMH, which can support users with visual or hearing impairments. 
\citeauthor{paneva20haptiread} use UMH to communicate Braille~\cite{paneva20haptiread} as an alternative to pin-array tactile displays~\cite{zeng25pinarray}.
With personalized rehabilitation programs in mind, \citeauthor{ragolia24rehabilitation} build a prototype to study how arm and hand stiffness influence tactile perception~\cite{ragolia24rehabilitation}. 
In the medical domain, rendering an acoustic force field more accurately aids ultrasound therapy in fracture-healing by avoiding ultrasound irradiation to an unaffected area~\cite{shinato23therapy}.
To this end, \citeauthor{shinato23therapy} develop an algorithm that can handle a higher resolution of focal points, especially those close-by~\cite{shinato23therapy}.

All these examples are little more than a glimpse into past and ongoing research, as extensive surveys and books on (ultrasound) mid-air haptics show~\cite{carlos21surveymobilear,georgiou2022ultrasound,qi2024bridging,kwok2024unobtrusive}.
However, the core concepts of focal point synthesis, perceptual resolution, and volumetric rendering underpin these applications.

\section{Bridging Disciplines for Human-Centred Innovation}
The evolution of ultrasound mid-air haptics highlights how substantial advances in HCI emerge when technical depth and human-centred aims are developed in parallel. 
Early advances in UMH required expertise far beyond traditional interaction design: applied mathematics and physics for modelling acoustic fields and solving linear systems, mechanical engineering for transducer design, and computer science for interactive rendering. 
These contributions transformed abstract acoustic fields into perceivable and expressive haptic experiences that support interaction, immersion, and accessibility, demonstrating what becomes possible when disciplinary boundaries are intentionally crossed.

As interactive technologies become increasingly ubiquitous and permeate everyday environments, HCI faces growing complexity in both its technical and societal responsibilities. 
The trajectory of UMH illustrates not only the benefits but also the necessity of bridging technical and human-centred domains. 
Perceptual research and interaction design rely on robust models and algorithms, while technical breakthroughs gain societal impact when guided by human-centred questions. 
By embracing interdisciplinary collaboration, particularly with applied mathematics, physics, and the machine learning community, HCI can more effectively transform sophisticated computational capabilities into interactive technologies that benefit diverse users and broaden who can participate in future digital experiences.

\bibliographystyle{ACM-Reference-Format}
\bibliography{sample-base}

\end{document}